\documentstyle[aps,preprint,prb]{revtex}
\tightenlines
\begin{document}
\def\vk{\vec k} 
\def\br{{\bf r}}
\title{\bf Impurity Doping Effect in High $T_{c}$ Superconductors}
\author{Yong-Jihn Kim and K. J. Chang}
\address{Department of Physics,  Korea Advanced Institute of Science and 
Technology,\\
 Taejon 305-701, Korea}
\maketitle
\begin{abstract}

It has been observed that impurity doping and/or ion-beam-induced
damage in high $T_{c}$ superconductors 
cause a metal-insulator transition and thereby suppress 
the critical temperature.
Based on our recent theory of the weak localization effect
on superconductivity, we examine the variation of
$T_{c}$ with increasing of impurity concentration $\rm (x)$ in
$\rm{La_{1.85}Sr_{0.15}Cu_{1-x}A_{x}O_{4}}$ systems, 
where A $=$ Fe, Co, Ni, Zn, or Ga.
We find that the doping impurity decreases the scattering matrix elements
for electron-electron attractions, such as
 $V_{nn'}=-V[1-{2\over \pi k_{F}\ell} ln(L/\ell)]$,
where $L$ and $\ell$ are the inelastic and elastic mean
free paths, respectively. Using the mean free path $\ell$ determined from 
resistivity data,  we find good agreements between our
calculated values for $T_{c}$ and experimental data except for Ni-doped 
samples, where  Ni impurities may enhance the pairing interaction.

\vspace{1pc}

\end{abstract}

\vfill\eject

Although impurity scattering may provide a clue to understanding the
superconducting  mechanism of high $T_{c}$ superconductors,
its effect is not well understood yet.$^{1-3}$
In strongly correlated systems such as high $T_{c}$ superconductors,
impurity potentials are significantly renormalized by
strong electron-electron interactions.$^{4}$ 
In particular, (unitarity) resonant impurity scattering seems to be important.
Using the marginal Fermi liquid theory,
Kotlier and Varma$^{4}$ showed that scattering by ordinary impurities
is close to the unitarity limit. 
Recently, it was argued that very low concentrations
of impurities in non-Fermi-liquid systems lead to a vanishing of density of states 
at the chemical potential and infinite resistivity for $T\rightarrow 0$.$^{5}$
In a one-channel Luttinger liquid with repulsive interactions, 
Kane and Fisher$^{6}$ showed that electrons are completely reflected even by 
the smallest scatterer as if it were effectively infinite. 
Hirschfeld and Goldenfeld$^{7}$ considered the effect of resonant
scattering on the penetration depth of a $d$-wave superconductor.
Poilblanc, Scalapino and Hanke$^{8}$ showed by 
an exact diagonalization technique for small clusters that strong correlations
give rise to resonant scattering due to local defects in the $t-J$ model.
In addition, Ziegler and his coworkers$^{9}$ introduced a simple static 
approximation to the effective impurity potential which reproduces
resonant scattering for both two-dimensional $t-J$ and Hubbard models.
Nevertheless, the impuirty effect on the transition tempearture of high $T_{c}$ 
superconductors remains to be unsettled, because
the effect of resonant scattering on the superconducting temperature
is not much different from that of Born scattering.$^{3, 10}$

We note that experimental results show consistently that impurity 
doping and ion-beam-induced damage give rise to a metal-insulator 
transition and thereby suppress $T_{c}$.$^{11-15}$ 
Recently, we studied the weak localization effect on low $T_{c}$ 
superconductors,$^{16}$ and found that for
two-dimensional systems weak localization 
decreases effective interactions, similarly to conductivity.
Here we use the same theoretical approach to investigate the impurity doping 
effect on the superconducting temperature for $\rm La_{1.85}Sr_{0.15}CuO_{4}$ 
with Cu replaced with 3d transition elements or Ga. 
The effect of strong correlation is taken into account by
the simple renormalization of the impurity potential.$^{9}$ The mean free path 
$\ell$  is determined from resistivity data.
We find good agreements between our theoretical calculations and
experimental data except for Ni-doped samples.
In Ni-doped case, the suppression of $T_{c}$ is much slower than the theory 
predicts, implying that Ni may enhance the pairing interaction.

Let us consider a simple pairing potential, $V(|{\br}_{1}-{\br}_{2}|)$,
between electrons of a pair.
The matrix elements of $V(r)$ for the plane wave states are 
\begin{eqnarray}
V_{\vk\vk'}&=&\int\int d{\br}_{1}d{\br}_{2} e^{-i\vk'\cdot ({\br}_{1}-{\br}_{2})}
V(|{\br}_{1}-{\br}_{2}|)e^{i\vk\cdot ({\br}_{1}-{\br}_{2})}\nonumber\\
&\cong& V_{o}+V_{1}{\hat k}\cdot{\hat k}'+
{V_{2}\over 2}\lbrack 3({\hat k}\cdot{\hat k'})^{2}-1\rbrack
+ \cdots.
\end{eqnarray}
For a $s$-wave pairing, $V_{\vk\vk'}$ is assumed to be a constant $V_{0}<0$,
whereas for a $d$-wave pairing, $V_{\vk\vk'}$ depends
on the unit vectors $\hat k$ and $\hat k'$, such as 
\begin{equation}
V_{\vk\vk'}={V_{2}\over 2}\lbrack 3({\hat k}\cdot{\hat k'})^{2}-1\rbrack.
\end{equation}
In the presence of impurities, the pairing interaction
between scattered basis pairs $(\psi_{n},\psi_{\bar n})$ and
$(\psi_{n'},\psi_{\bar n'})$ is given by$^{17}$
\begin{equation}
V_{nn'}=\int\int d{\br}_{1}d{\br}_{2} \psi_{n'}^{*}({\br}_{1})\psi_{\bar n'}^{*}
({\br}_{2}) V(|{\br}_{1}-{\br}_{2}|)
\psi_{\bar n}({\br}_{2})\psi_{n}({\br}_{1}),
 \end{equation}
where $\psi_{\bar n}$ denotes the time-reversed partner of $\psi_{n}$.
The metal-insulator transition driven by the impurity doping may be
understood by the localization of wave function.$^{18,19}$ In  
high $T_{c}$ superconductors, resistivity data showed that correlation 
enhances significantly the strength of impurity potential. 
Kaveh and Mott$^{19}$ derived the weakly localized scattered states
in the form of power-law and extended wave functions for two-dimensional 
systems,
\begin{equation}
\psi_{n}({\br}) = A_{2}e^{i\vk\cdot{\br}} + B_{2}{e^{ikr}\over r},
 \end{equation}
\begin{equation}
A_{2}^{2}=1-2\pi B_{2}^{2}ln(L/\ell), \ \ B_{2}^{2}={1\over \pi^{2}k_{F}\ell},
 \end{equation}
where $\ell$ and $L$ are the elastic and inelastic mean free paths, respectively.
We can calculate $V_{nn'}$ by substituting Eq. (4) into Eq. (3).
Since the Cooper pairs in a BCS condensate form bound states,
only the power-law wave functions within the BCS coherence length are 
relevant.$^{20}$
Thus, the Cooper pair wave functions basically consist of the
plane waves with reduced amplitudes.
The resulting matrix elements are$^{21,22}$
\begin{equation}
V_{nn'}=V_{\vk\vk'}\lbrack 1-{2\over \pi k_{F}\ell}ln(L/\ell)\rbrack,
 \end{equation}
for both the $s$- and $d$-wave pairings.
In the $d$-wave pairing, Eq. (6) does not include the effect of impurity
scattering in the Borm limit. Then,
the effective coupling constant $\lambda_{eff}$ becomes
\begin{equation}
\lambda_{eff}=\lambda \lbrack 1-{2\over \pi k_{F}\ell}ln(L/\ell)\rbrack,
 \end{equation}
which satisfies the modified BCS $T_{c}$ equation,
\begin{equation}
T_{c}=1.13\epsilon_{c} exp(-1/\lambda_{eff}).
 \end{equation}

Xiao and his coworkers$^{11}$ conducted a systematic study on the effect of Cu-site
doping in  $\rm La_{1.85}Sr_{0.15}CuO_{4}$ polycrystals and determined
the relation between $T_{c}$ and dopant level  
for  Fe-, Co-, Ni-, Zn-, Ga-, and Al-impurities.
Cieplak and his coworkers$^{12}$  also measured 
resistances for the same materials with different impurities 
(Fe, Co, Ni, Zn, and Ga) down to 50 mK and found the metal-insulator
transitions produced by impurities.  
Here we use Eq. (8) to calculate $T_{c}$ as a function of dopant 
concentration,
assuming that the cutoff energy $\epsilon_{c}=450\rm{K}$ and the Fermi wave vector
$k_{F}=0.2\rm{\AA}^{-1}$.
In this case, $\lambda$ is chosen to give $T_{co}=37\rm{K}$ without impurities.
The inelastic mean free path is obtained from disordered
two-dimensional systems, which is employed for all the systems considered here,
 with the $1/\sqrt{lnT}$ 
dependence removed, i.e, $L=500\rm{\AA}/\sqrt{T}$.$^{18}$ Because of the Sr 
doping, the elastic mean free path is estimated to be about $50\rm{\AA}$ without 
Cu-site doping.$^{13}$
Then the total elastic mean free path $\ell_{tot}$ (in unit of $\rm{\AA}$) is 
\begin{equation}
{1\over \ell_{tot}}={1\over 50 }+{1\over \ell_{imp}},
\end{equation}
where
$\ell_{imp}$ is the mean free path caused by impurity doping for Cu.
For two-dimensional layered systems such as high $T_{c}$ superconductors,
the mean free path $\ell_{imp}$  can be obtained from the Drude formula, 
$k_{F}\ell=h/R_{\Box} e^{2}$, where $R_{\Box}$ is the sheet resistance,
defined as $R_{\Box}=\rho/d$.
In this case, $d$ is the interlayer distance (6.5 $\rm{\AA}$ in the LSCO system)$^{23}$ 
and $\rho$ is resistivity.
From experimental data by Cieplak et al.,$^{12}$ the resistivities of Ga-,
Zn-, Fe-, Co-, and Ni-doped systems can be fitted  such as 
$\rho_{Ga}= 198.41\times c$,
$\rho_{Zn}=147.11\times c + 40$,   
$\rho_{Fe}=227.4\times c + 33.7$,   
$\rho_{Co}=190.67\times c $,  and 
$\rho_{Ni}=164.35\times c +12$, respectively, where
$c$ denotes the concentration of dopants. 

The superconducting transition temperatures are plotted  as 
a function of dopant concentration and compared with experiments in Figs. 1 and
2. For nonmagnetic (Ga and Zn) and magnetic (Fe and Co) 
impurities, the agreements between theory and experiment are satisfactory,
considering the simple approximation of strong correlation effect.
It is clearly seen that the impurity doping gives rise
to a metal-insulator transition and  suppresses $T_{c}$. We find that 
only the strength of  (effective) impurity potential is important and 
the effect of nonmagnetic impurities on destroying superconductivity
is  similar to magnetic impurities, which agrees well with experiments.
Although the Zn and Ga atoms have filled d-shells whereas the
Fe, Co, and Ni atoms have unfilled d-shells,  
it appears that the variation of $T_{c}$ with impurity doping does not
depend significantly on the valence states of impurities. We find that 
the initial drops of $T_{c}$ for all the samples are larger than  
the theory predicts.  

Fig. 3 shows the variation of $T_{c}$ for Ni-doped samples. 
We find a large deviation between the theoretical and experimental
curves. Although the origin of the discrepancy is not clear,
it seems that Ni may enhance the pairing interaction.
It is interesting to note that a Ni impurity in $\rm {YBa_{2}Cu_{3}O_{7-x}}$
has an unpaired spin of $S={1\over 2}$ rahter than $S=1$, as expected
for $\rm Ni^{2+}$.$^{24}$
It is also possible  that Ni-doped samples may develop some  
inhomogeneous granular structures$^{25}$ or stripe phases$^{26}$ and 
their superconducting phases may  persist for very high impurity concentrations.
However, x-ray diffraction measurements in Ni-doped single-crystals 
$\rm YBa_{2}Cu_{3}O_{7-x}$ did not support this idea.$^{27}$ 

In the underdoped regime, experiments showed that impurity scattering 
produces a large residual resistivity close to (or even 
larger than) the unitarity limit and the metal-insulator transition 
occurs almost simultaneously when $T_{c}$ drops to zero.$^{13,14}$
Consequently, our theory which considered the optimum doping regime
is also applicable for the underdoped regime.
As the hole density increases from the optimum doping, $T_{c}$ drops
to zero more quickly before the metal-insulator transition is reached.
This behavior may be related to the dimensional crossover from 
two-dimension to three-dimension. 
Nevertheless, our theory may explain the decrease of $T_{c}$ in 
moderately overdoped samples. 
Recently, Suryanarayanan and his coworkers$^{28}$ found the recovery of superconductivity
in $\rm Y_{1-x}Ca_{x}SrBaCu_{2.6}Al_{0.4}O_{6+z}$ when Y is substituted by
Ca. This result may be understood if we consider the importance of
the metal-insulator transition caused by impurity doping.
Since Ca increases the hole density in the Cu-O planes,
the mobility edge is shifted with respect to the Fermi energy, 
so that the system becomes metallic and superconducting. 
The transition between the insulating and superconducting phases was also
explored in thin films such as $\rm DyBa_{2}Cu_{3}O_{7}$,$^{29}$ 
$\rm Nd_{2-x}Ce_{x}CuO_{4}$,$^{30}$ and $\rm YBa_{2}Cu_{3}O_{7}$,$^{31}$ 
where the two-dimensional localization 
is the main reason for the suppression of  superconductivity and 
the critical sheet resistance lies in the range of 6 to 8 $\rm k\Omega$.
In $\rm YBa_{2}Cu_{3}O_{7-x}/PrBa_{2}Cu_{3}O_{7-x}$ superlattices,$^{32,33}$
the decrease of $T_{c}$ in isolated $\rm YBa_{2} Cu_{3}O_{7-x}$
layers with increasing sheet resistance is on the 
scale of $\hbar/e^{2}$, as observed for ordinary superconductors.$^{32}$
This behavior may also be explained in terms of  our theory.

In conclusion, we have studied the impurity doping effect on the 
superconducting temperatures of $\rm La_{1.85}Sr_{0.15}Cu_{1-x}A_{x}O_{4}$ 
systems (A=Fe, Co, Ni, Zn, and Ga)
based on our recent theory of weak localization effect on superconductors.
The strong correlation effect is taken into account by a simple 
renormalization of the effective impurity potential.
The calculated values for $T_{c}$ are in good agreement with
experiments, while 
in Ni-doped samples the agreement was poor, implying that Ni may
enhance the pairing interaction.
\vspace{1pc}

\centerline{\bf ACKNOWLEDGMENTS}

We are grateful to Yunkyu Bang for discussions.
This work is supported by the Brain pool project of KOSEF and the MOST.

\vfill\eject
\centerline{\bf Figure Captions}
\vspace{1pc}
\noindent
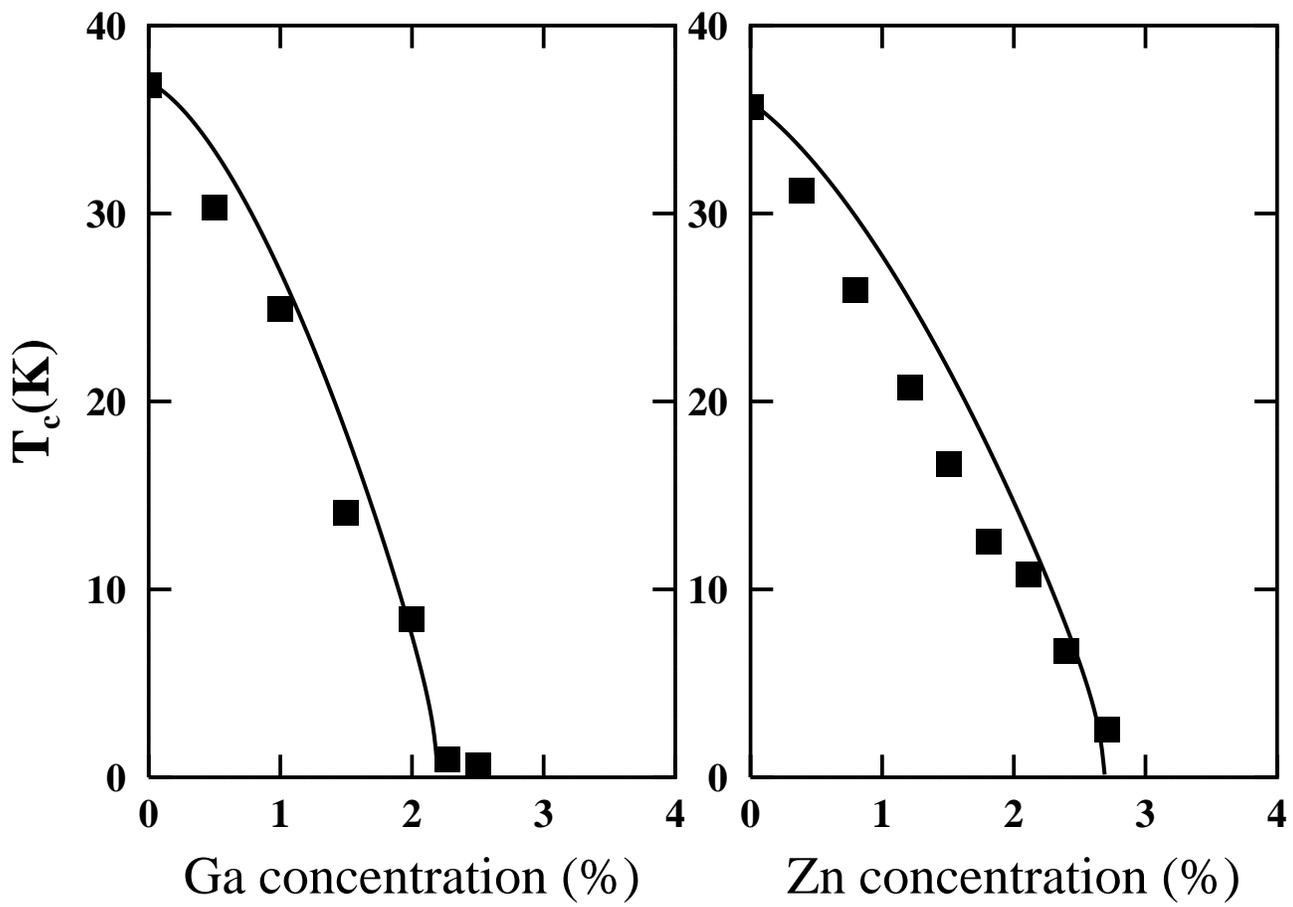
{\bf Fig. 1} Variation of $T_{c}$ with dopant concentration 
for $\rm La_{1.85}Sr_{0.15}Cu_{1-x}A_{x}O_{4}$ 

(A=Ga and Zn). Experimental data are from Ref. 11. 

\noindent
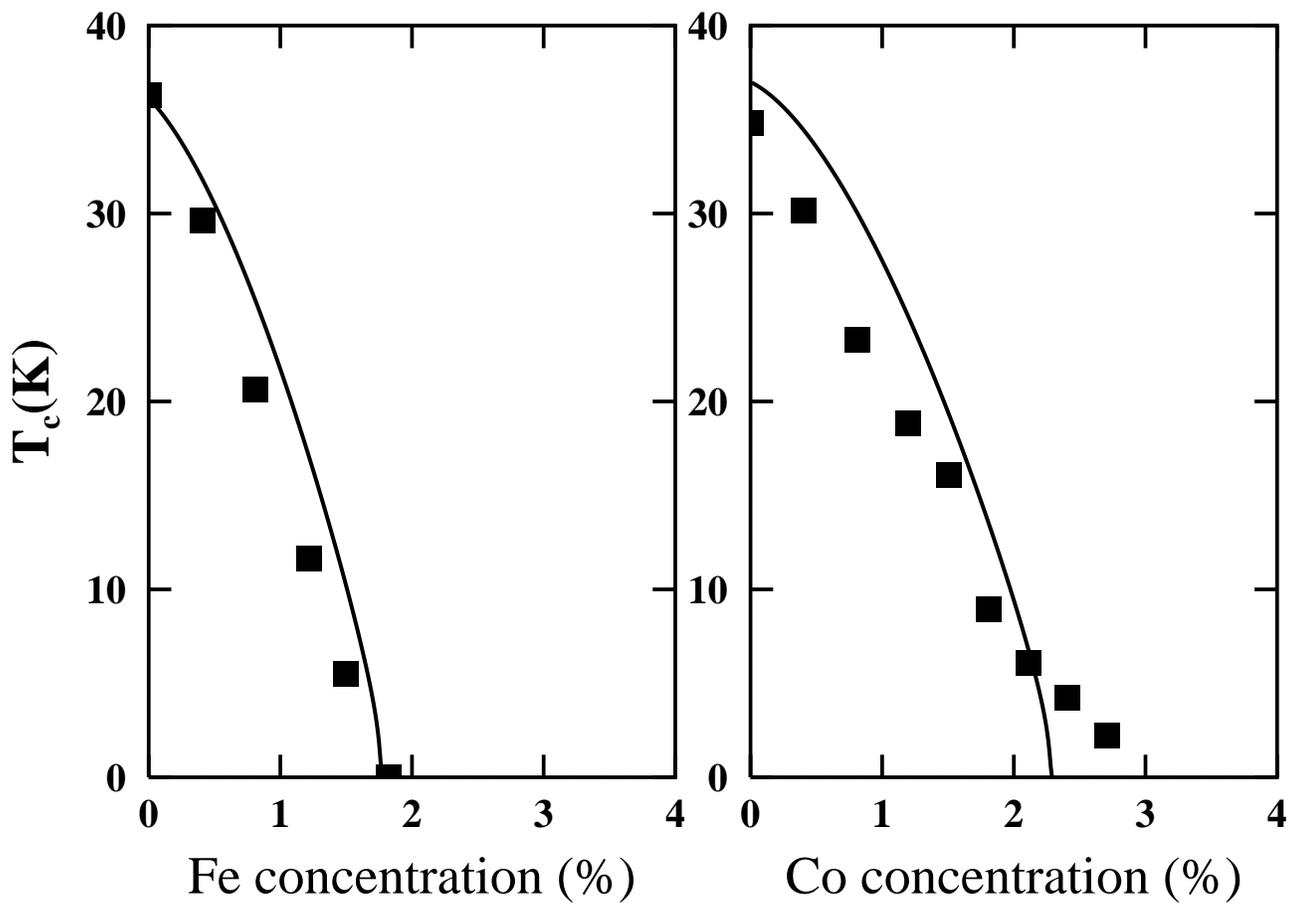
{\bf Fig. 2} Variation of $T_{c}$ with magnetic dopant content
for $\rm La_{1.85}Sr_{0.15}Cu_{1-x}A_{x}O_{4}$ 

(A=Fe and Co).  Experimental data are from Ref. 11. 

\noindent
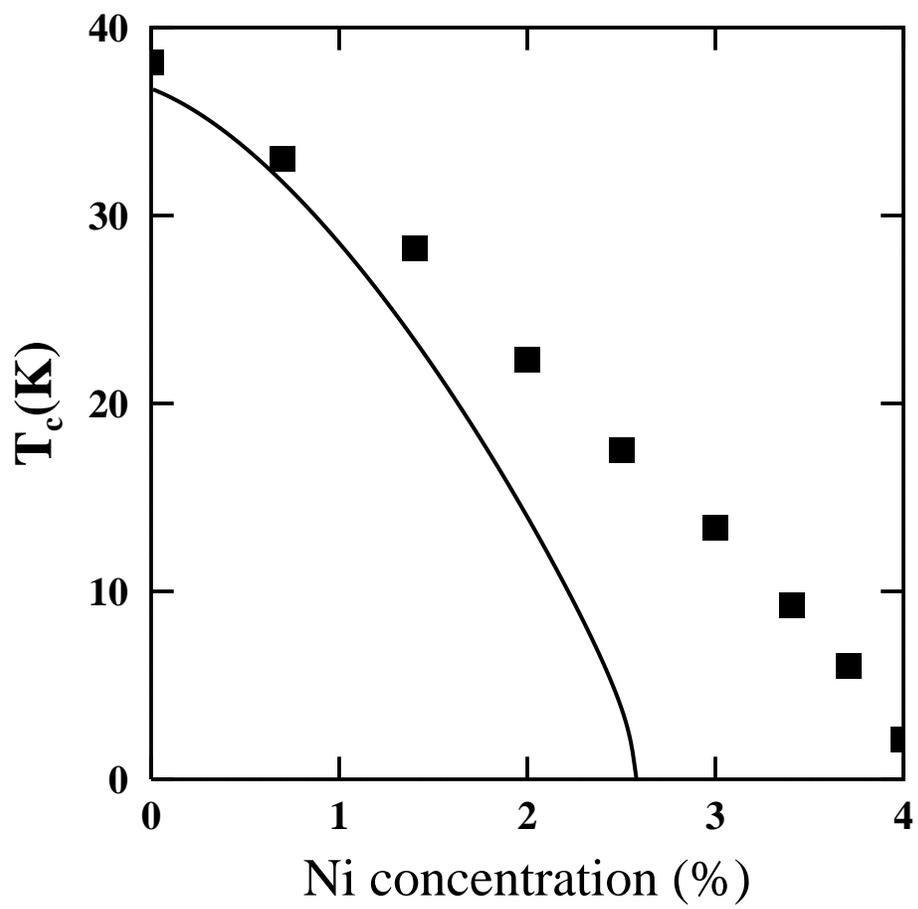
{\bf Fig. 3} Variation of $T_{c}$ with Ni concentration
for $\rm La_{1.85}Sr_{0.15}Cu_{1-x}Ni_{x}O_{4}$.
 Experimental 

data are from Ref. 11. 


\begin{references}
\bibitem{1} R. J. Radtke, K. Levin, H.-B. Schuttler, and M. R. Norman, Phys. Rev. B {\bf 48},
                       653 (1993). 
\bibitem{2} Y.-J. Kim and K. J. Chang, J. Korean Phys. Soc. {\bf 31}, S298 (1997).
\bibitem{3} M.-A. Park, M. H. Lee, and Y.-J. Kim, Mod. Phys. Lett. B {\bf 11}, 719 (1997).
\bibitem{4} G. Kotlier and C. M. Varma, Physica A {\bf 167}, 288 (1990).
\bibitem{5} C. M. Varma, preprint (1997).
\bibitem{6} C. L. Kane and M. P. A. Fisher, Phys. Rev. Lett. {\bf 68}, 1220 (1992).
\bibitem{7} P. J. Hirshfeld and N. Goldenfeld, Phys. Rev. B {\bf 48}, 4219 (1993). 
\bibitem{8} D. Poilblanc, D. J. Scalapino, and W. Hanke, Phys. Rev. Lett.
{\bf 72}, 884 (1994).
\bibitem{9} W. Ziegler, D. Poilblanc, R. Preuss, W. Hanke, and D. J. Scalapino,
Phys. Rev. B {\bf 53}, 

8704 (1996).
\bibitem{10} P. Arberg and J. P. Carbotte, Phys. Rev. B {\bf 50}, 3250 (1994). 
\bibitem{11} G. Xiao, M. Z. Cieplak, J. Q. Xiao, and C. L. Chien, Phys. Rev. 
   B {\bf 42}, 8752 (1990). 
\bibitem{12} M. Z. Cieplak, S. Guha, H. Kojima, P. Lindenfeld,
          G. Xiao, J. Q. Xiao, and C. L. Chien, 

Phys. Rev. B {\bf 46}, 5536 (1992). 
\bibitem{13} V. P. S. A. Awanda, S. K. Agarwal, M. P. Das, and A. V. Narlikar, 
J. Phys.: Condens. 

Matter {\bf 4}, 4971 (1992).
\bibitem{14} Y. Fukuzumi, K. Mizuhashi, K. Takenaka, and S. Uchida, Phys.
    Rev. Lett. {\bf 76}, 684 (1996). 
\bibitem{15} J. M. Valles, Jr., A. E. White, K. T. Short, R. C. Dynes,
          J. P. Garno, A. F. J. Levi, M.

Anzlowar, and K. Baldwin, Phys. Rev.
         B {\bf 39}, 11599 (1989). 
\bibitem{16} Y.-J. Kim and K. J. Chang, unpublished (1997).
\bibitem{17} P. W. Anderson, J. Phys. Chem. Solids, {\bf 11}, 25 (1959).
\bibitem{18} P. A. Lee and T. V. Ramakrishnan, Rev. Mod. Phys. {\bf 52},
               287 (1985).
\bibitem{19} N. F. Mott and M. Kaveh, Adv. Phys. {\bf 34}, 329 (1985).
\bibitem{20} Y.-J. Kim and A. W. Overhauser, Phys. Rev. B {\bf 49}, 15799 
     (1994).
\bibitem{21} Y.-J. Kim, Mod. Phys. Lett. B {\bf 10}, 555 (1996).
\bibitem{22} Y.-J. Kim, Int. J. Mod. Phys. Lett. B {\bf 11}, 1731 (1997).
\bibitem{23} G. S. Boebinger, Y. Ando, A. Passner, T. Kimura, M. Okuya, 
      J. Shimoyama, K. Kishio, 

K. Tamasaku, N. Ichikawa, and S. Uchida, Phys. Rev. Lett. {\bf 77}, 5417 (1996).
\bibitem{24} P. Mendels et al., Physica 235-240, 1595 (1994).
\bibitem{25} B. I. Belevtsev, Sov. Phys. Usp. {\bf 33}, 36 (1990).
\bibitem{26} J. M. Tranquada, J. D. Axe, N. Ichikawa, A. R. Moodenbaugh,
                P. K. Mukopadhyay, and 

H. Pankowska, Sol. Sta. Commun.   {\bf 81}, 593 (1992).
\bibitem{27} S. A. Hoffman, M. A. Castro, G. C. Follis, and S. M. Durbin,
         Phys. Pev. B {\bf 49}, 12170 

(1993).
\bibitem{28} R. Surynarayanan, L. Ouhammou, M. S. R. Rao, O. Gorochov,
                P. K. Mukopadhyay, and H. Pankowska, Sol. Sta. Commun. 
              {\bf 81}, 593 (1992).
\bibitem{29} T. Wang, K. M. Beauchamp, D. D. Berkley, B. R. Johnson, J.-X. Liu,
            J. Zhang, and A. 

M. Goldman, Phys. Rev. B {\bf 43}, 8623 (1991).
\bibitem{30} S. Tanda, M. Honma, and T. Nakayama, Phys. Rev. B {\bf 43}, 
          8725 (1991). 
\bibitem{31} T. Terashima, K. Shimura, Y. Bando, Y. Matsuda, A. Fujiyama,
            and S. Komiyama, Phys. 

Rev. Lett. {\bf 67}, 1362 (1991).
\bibitem{32} J.-M. Triscone, \O. Fischer, O. Brunner, L. Antognazza, A. D.
   Kent, and M. G. Karkut, 

Phys. Rev. Lett. {\bf 64}, 804 (1990). 
\bibitem{33} Q. Li, X. X. Xi, X. D. Wu, A. Inam, S. Vadlamannati, W. L. McLean,
        T. Venkatesan, R. 

Ramesh, D. M. Hwang, J. A. Martinez, and L. Nazar,
           Phys. Rev. Lett. {\bf 64}, 3086 (1990).
\end{references}
\end{document}